\documentclass[article]{revtex4}

\usepackage{graphicx}
\usepackage{bm}
\usepackage{amsmath,amssymb}
\usepackage{siunitx}

\begin{document}

\title[Flow field from transient bubble oscillation in a narrow gap]{Flow field from transient bubble oscillation in a narrow gap: numerical simulations and effect on biological cells}

\author{Milad Mohammadzadeh}
\affiliation{Division of Physics and Applied Physics, School of Physical and Mathematical Sciences, Nanyang Technological University, Singapore 637371}

\author{Fenfang Li}
\affiliation{Division of Physics and Applied Physics, School of Physical and Mathematical Sciences, Nanyang Technological University, Singapore 637371}

\author{Claus-Dieter Ohl}
\email{cdohl@ntu.edu.sg}
\affiliation{Division of Physics and Applied Physics, School of Physical and Mathematical Sciences, Nanyang Technological University, Singapore 637371}

\begin{abstract}
The flow driven by a rapidly expanding and collapsing cavitation bubble in a narrow cylindrical gap is studied with the volume of fluid method. The simulations reveal a developing plug flow during the early expansion followed by flow reversal at later stages. An adverse pressure gradient leads to boundary layer separation and flow reversal, causing large shear stress near the boundaries. Analytical solution to a planar pulsating flow shows qualitative agreement with the CFD results. The shear stress close to boundaries has implications to deformable objects located near the bubble: experiments reveal that thin, flat biological cells entrained in the boundary layer become stretched, while cells with a larger cross-section are mainly transported with the flow. 
\end{abstract}

\maketitle

\section{Introduction}\label{sec:introduction}
Strongly oscillating bubbles in narrow gaps are commonly found in microfluidic applications of cavitation. These bubbles can be generated with a focused laser pulse \cite{chen2006interaction, zwaan2007controlled, pedro08}, acoustically excited capillary waves \cite{tandiono2010capillary}, or through spark discharges \cite{azam2013dynamics}. Applications of these transient pulsating flows span cell stretching \cite{quinto2011red, li2013yield}, liquid pumping \cite{dijkink2008laser}, switching and sorting \cite{Wu08, Wu12}, mixing \cite{Hellman07}, and droplet generation \cite{Park11}.

Modelling the fluid flow in these applications has been done to various degrees of sophistication. The problem can be simplified as a planar inviscid flow leading to a Rayleigh type equation in cylindrical coordinates \cite{Pedro09}. This potential flow description has been extended to non-spherical bubbles in narrow gaps \cite{Lim10}. However, a notable deficiency of these approximations is their inability to model boundary layers, which are important when dealing with suspended objects near the walls, e.g. flat red blood cells and thin elastic objects such as nanowires \cite{Pedro10}.  Our recent experiments on bubble-induced cell stretching, such as red blood cells \cite{li2013yield}, gave motivation to model the fluid flow in order to understand the underlying flow patterns that causes cell deformation. In the present work, we focus on the 3-dimensional structure of the liquid flow, i.e. the formation of boundary layers during the expansion and collapse cycle of a single transient bubble.

In general, this confined flow may be simplified to an axisymmetric radial flow forced by a time-dependent source at the origin. Axisymmetric radial flows in narrow gaps have been studied experimentally and analytically in the past fifty years due to their relevance in industrial applications such as radial viscometers, radial diffusers, non-rotating air bearings, and disk type heat exchangers. For an oscillating source between two parallel plates, in which the source strength varies sinusoidally about a zero-mean value, \citet{elkouh1975oscillating} obtained an analytical solution and reported reversed flow near the walls. \citet{zitouni1997purely} report on analytical power series solution to purely accelerating and decelerating flows between two flat disks, which is later studied numerically in \citep{ghaly2001numerical} as well. Although the flow reversal is not captured with the solution provided in \citep{zitouni1997purely}, it can be deduced to occur once the derivative of the velocity in axial direction becomes zero at the wall. \citet{von1999note} completed the work of \citet{zitouni1997purely} by finding analytical solutions for the cases where flow reversal indeed happens, i.e. where the flow is neither purely accelerating nor decelerating.

Several groups have investigated bubble pulsations in a confinement. \citet{cui2006bubble} studied analytically the response of an acoustically driven spherical bubble confined between two parallel plates. In this study, although the bubble is much smaller than the gap height, the channel walls affect the bubble dynamics. As the spherical bubble is confined between two plates, decreasing the channel height reduces the resonance frequency and the maximum response amplitude of the bubble. Large bubbles, in contrast, obtain a cylindrical shape bounded by the walls, which is referred to as a cylindrical bubble. \citet{ilinskii2012models} obtained a solution for harmonic cylindrical bubble pulsations in infinite and compressible liquid. They compared their model with the Gilmore equation for cylindrical bubble oscillations and also with the commonly used 2-dimensional Rayleigh-Plesset equation, emphasizing on the role of liquid compressibility.

A more detailed study of the liquid flow field induced by bubble activity in a confinement has been reported by \citet{ye2004direct}. They have conducted direct numerical simulation of microbubble expansion and shrinkage in a long tube, which represents the bubble activity in human vasculatur system during gas embolotherapy. An improved model with flexible walls was presented in \citep{ye2006microbubble}.

In a second application of microbubble expansion, the deformation of cells has been modelled with a boundary element method by \citet{tandiono2013resonant} which accounts for the membrane tension of the cell. Their finding is that deformation of a cell, modelled as a liquid droplet, is maximised if the resonance frequency of its surface mode matches the oscillation period of the bubble. This shape frequency is dependent on the density contrast between the liquids and the membrane tension. The model provides a physical explanation for why the shape of an elastic object in a symmetric back-and-forth motion does not return to its original state, in contrast to a fluid particle in a homogeneous flow. Here, however, the no-slip boundaries were ignored.

In the present work we focus on unsteady boundary layers generated in close proximity of an oscillating bubble between two parallel disks using numerical simulations of the flow. The rapidly expanding bubble is assumed to be created by an intense, focused laser pulse as a method of impulsive deposition of energy in the liquid. The simulation results are compared to an analytical expression for planar flow induced by an oscillating pressure gradient. A simple experiment using deformable biological cells of various sizes emphasizes the importance of the boundary layer, and the flow field obtained from simulations increases our understanding about the deformation of elastic objects in proximity of a confined cavitation bubble. 

\section{Numerical simulation}\label{sec:numerical_simulation}
\subsection{Computational domain}\label{subsec:cfd_setup}

\begin{figure}
\begin{center}
\includegraphics[width=0.6\textwidth]{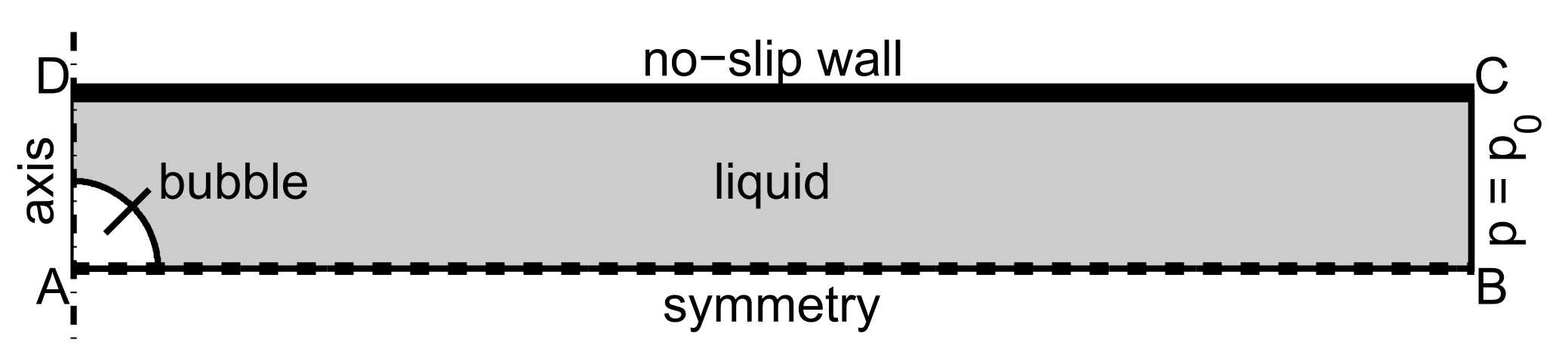}
\end{center}
\caption{Sketch of the computational domain using axisymmetry and symmetry with respect to the centre-plane of the gap.}
\label{fig:computational_domain}
\end{figure}

The problem being modelled is the expansion and shrinkage of an initially spherical bubble, created by a focused, high-power laser pulse. The bubble is located at the centre and between two parallel discs separated by a gap of height {$h=$\SI{20}{\micro m}}.  We assume axisymmetry and only model the upper quarter of the gap, i.e. we utilise symmetry to reduce the computational cost, see figure \ref{fig:computational_domain}. The computational domain is drawn to scale, with the length being {$\overline{\mbox{AB}}=$\SI{80}{\micro m}} and the hight {$h/2=\overline{\mbox{DA}}=$\SI{10}{\micro m}}.

The complex physical process of laser-matter interaction \citep{lauterborn2013shock} leading to a rapidly expanding bubble is greatly simplified by starting the simulation with a bubble of finite size, filled with non-condensible gas at high pressure, similar to previous work done by \citep{gonzalez2011cavitation} and \citep{tandiono2013resonant}. This simplification allows us to focus on fluid motion and avoid the complexity of propagation of shock waves and liquid compressibility. The liquid motion around the bubble happens in a time scale of tens of microseconds, while the acoustic transients are significant for at most hundreds of nanoseconds \citep{lauterborn2013shock, vogel1996shock}. Therefore, for the purpose of resolving the pressure and velocity field in the liquid surrounding the bubble, liquid compressibility effects and acoustic transients could be safely neglected.

\subsection{Numerical solver specifications}\label{subsec:solver_specifications}

The multiphase flow problem of the compressible gas and the incompressible liquid is modelled with volume of fluid (VOF) method accounting for interfacial tension but neglecting body forces using ANSYS Fluent 14.0 \citep{ansys201114}. The boundary conditions as depicted in figure \ref{fig:computational_domain} are along $\overline{\mbox{DA}}$ axis of symmetry, between $\overline{\mbox{AB}}$ symmetry, at $\overline{\mbox{BC}}$ constant pressure $p_0$, and no slip at $\overline{\mbox{CD}}$. In the VOF method a single set of momentum equations is solved for all phases, meaning that the pressure and velocity field are shared among all present phases, and the volume fraction of each phase is tracked throughout the domain. The momentum equation can be expressed as

\begin{equation}\label{eq:momentum}
  \frac{\partial}{\partial t} (\rho \vec{v})+\nabla . \left( \rho \vec{v} \vec{v} \right) = -\nabla p + \nabla . \left( \mu \left( \nabla \vec{v} + \nabla \vec{v}^T \right)  \right) + \vec{F}
 \quad,
\end{equation}

\noindent where $\rho$ is the density of the phases, $\vec{v}$ is the velocity, $p$ the pressure, and $\vec{F}$ is the surface force. The surface force is modelled as a continuum surface force (CSF) as proposed in \citep{brackbill1992continuum}. In ANSYS Fluent, the surface curvature is calculated from local gradients in the surface normal at the interface of phases. In solving the governing equations, the material properties, such as density, $\rho$, or viscosity, $\mu$, are calculated as volume-fraction-averaged properties.

Interface tracking is done by coupling the volume of fluid with the level-set method \citep{osher1988fronts,sussman1994level}. This allows for accurate interface tracking as well as mass conservation, in spite of the large density difference between the bubble content and the liquid.

In the present simulation, the pressure-implicit with splitting of operators (PISO) scheme is used for pressure-velocity coupling. The pressure staggering option (PRESTO!) is chosen for spatial discretisation of pressure while second order upwind differencing is used in solving the governing equations. The Geo-Reconstruct method is implemented for discretisation of volume fraction and interface reconstruction. An absolute convergence criteria of $10^{-6}$ is used for all governing equations and time step size of the simulation is {\SI{1e-8}{s}}. To assure the solution is mesh-independent, the simulations were conducted with two different grids, $5000$ and $80000$ elements, with an average of $0.4$ and {\SI{0.1}{\micro m}} element size respectively. The difference of the solutions obtained by the two grids is negligible, as presented in the evolution of bubble radius in figure \ref{fig:comparison}, therefore grid independence is indeed obtained. The results reported in this manuscript are for {\SI{0.1}{\micro m}} element size, while the mesh is refined for regions of high gradient, such as initial bubble-liquid interface and the vicinity of channel wall. The simulation time on a 2.1 GHz Intel Core i7 personal computer with 8 GB of RAM is approximately 24 hours for a single cycle of bubble expansion and shrinkage.

The initial pressure in the gas bubble of {$R(t=0)=$\SI{5}{\micro m}} is {\SI{100}{bar}}. The liquid is initially at atmospheric pressure {$p_0=$\SI{1}{bar}}. The liquid is water with a density of {$\rho_l=$\SI{998.2}{kg/m^3}} and a dynamic viscosity of {$\mu_l=$\SI{1e-3}{Pa s}}, while the gas viscosity is {$\mu_b=$\SI{1.34e-5}{Pa s}}, corresponding to water vapour. The ideal gas law is used for calculation of density in the compressible bubble content. This density calculation requires the solution to the energy equation \citep{ansys201114}, which is shared among both phases, similar to the momentum equation (\ref{eq:momentum}). The temperature field is initially assumed to be uniform in the computational domain and at {\SI{300}{K}}. The interfacial tension coefficient is {$\gamma=$\SI{7.2e-2}{N/m}}.

\section{CFD results and analysis of the flow field}\label{sec:cfd_results}

\subsection{Bubble evolution and liquid velocity profile}\label{subsec:bubble_evolution}

\begin{figure}
\begin{center}
\includegraphics[width=0.9\textwidth]{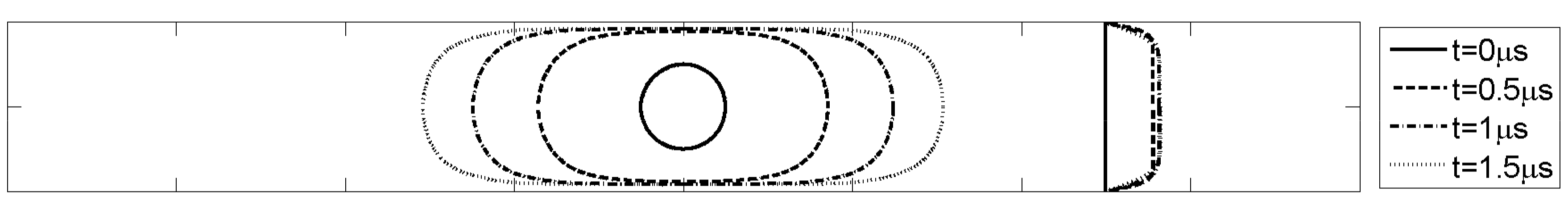}
\includegraphics[width=0.9\textwidth]{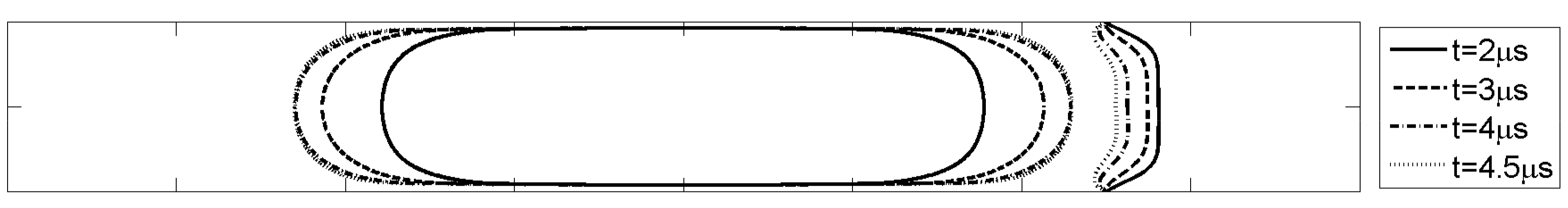}
\includegraphics[width=0.9\textwidth]{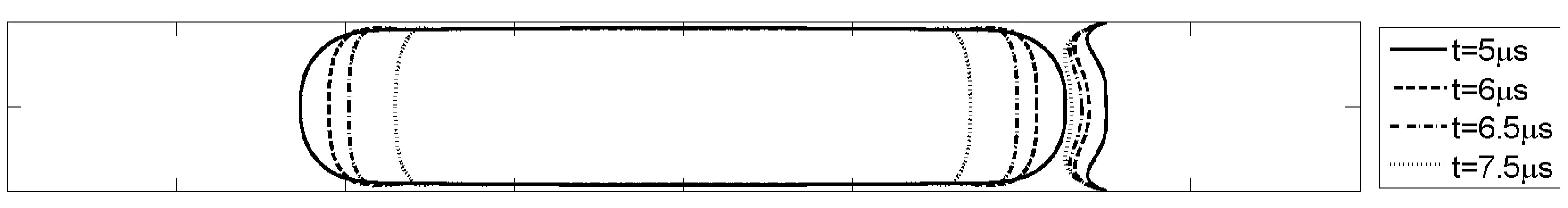}
\includegraphics[width=0.9\textwidth]{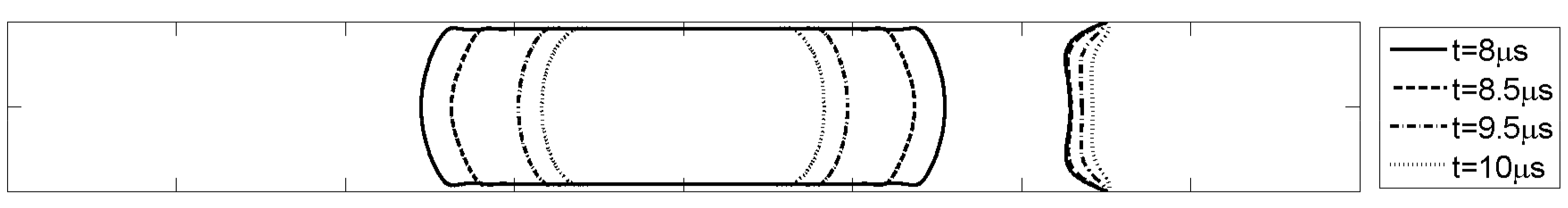}
\end{center}
\caption{CFD results of the bubble shape evolution together with the radial liquid velocity profile at {$r=$\SI{50}{\micro m}} from the centre of the bubble. Please note the reversal of the flow direction near the boundaries occurring before the flow in the centre of the gap.}
\label{fig:bubble_evolution}
\end{figure}

\begin{figure}
\begin{center}
\includegraphics[width=0.6\textwidth]{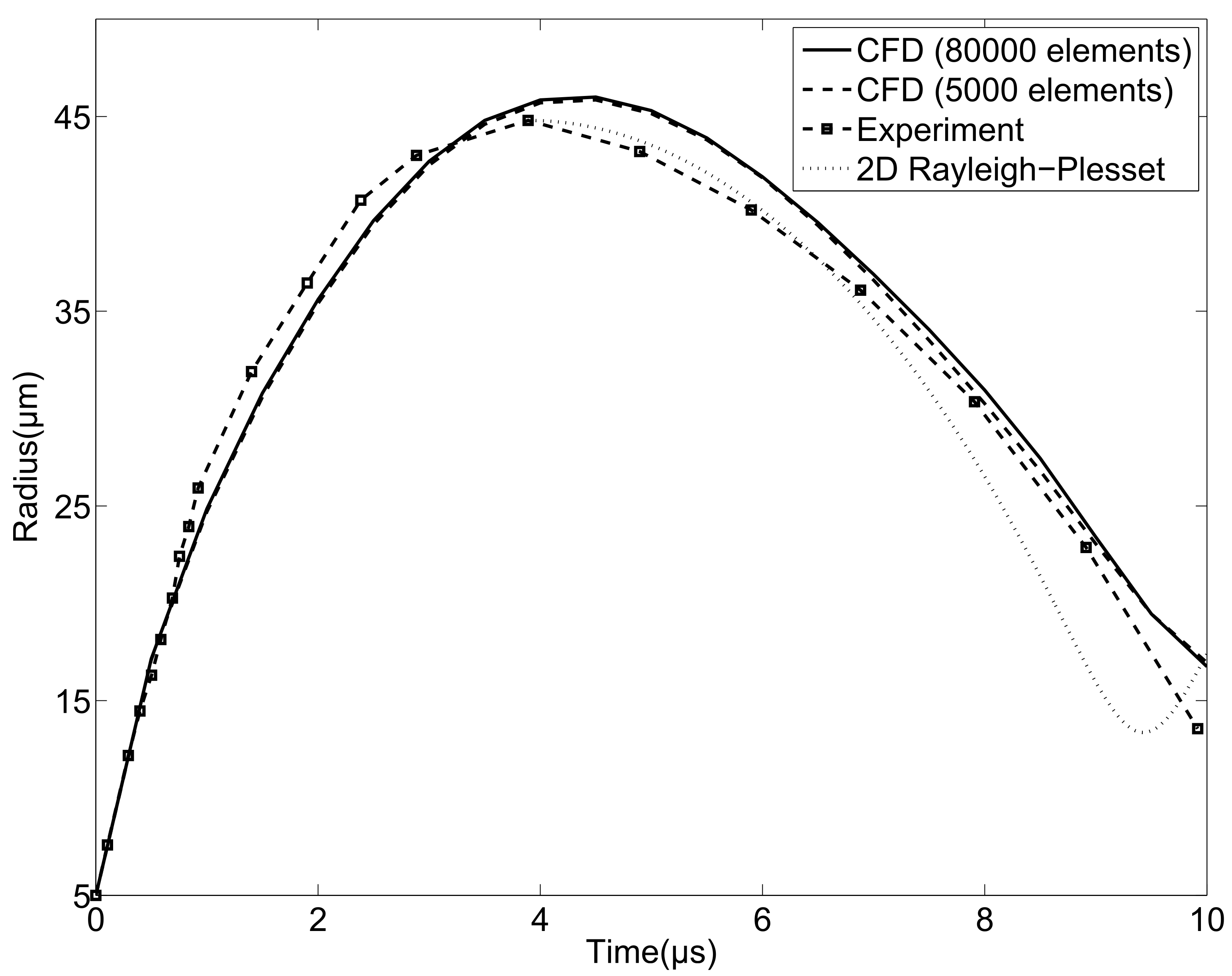}
\end{center}
\caption{Comparison of experimental cylindrical bubble dynamics \citep{zwaan2007controlled} (dashed line with squares) with the VOF solution (filled line and dashed line for two computational grids). Additionally, the solution to the 2D Rayleigh-Plesset equation starting from the maximum bubble radius is shown with a dotted line.}
\label{fig:comparison}
\end{figure}

Figure \ref{fig:bubble_evolution} depicts the temporal evolution of the bubble shape over a period of {\SI{10}{\micro s}} in a gap of {$h=$\SI{20}{\micro m}}. The spherical bubble quickly grows into a pancake shape, forming thin liquid films at the upper and lower solid walls. Maximum bubble radius is obtained after {\SI{4}{\micro s}}. During bubble shrinkage, the convex interface flattens and only increases curvature after {\SI{8}{\micro s}}; that is when the internal pressure increases and dampens the collapse of the bubble. The bubble collapses to its minimum volume at {\SI{10}{\micro s}} and rebounds afterwards (not shown here). Right to the bubble profile in figure \ref{fig:bubble_evolution}, the radial velocity profile at a distance of {$r=$\SI{50}{\micro m}} from the bubble centre is shown. Initially, the liquid is at rest and rapidly develops into a plug flow with strong wall shear stress. Gradually, a more parabolic profile develops. At the later expansion stage, the velocity near the walls is reduced and even reversed in direction, while the liquid velocity in the centre of the gap is still outwards and positive. This flow reversal near the boundaries will be discussed in detail below. The flow reversal becomes more pronounced during bubble collapse.

Figure \ref{fig:comparison} shows the projected bubble radius from figure \ref{fig:bubble_evolution} and compares it with the experimentally determined radii \citep{zwaan2007controlled} and a 2-dimensional Rayleigh-Plesset equation, e.g. see \citep{Pedro09}. For the chosen initial conditions, i.e. radius $R(t=0)$ and initial gas pressure, we find a good agreement between the VOF simulation and the experiment. In particular, the asymmetry of the bubble oscillation, having a faster expansion than the collapse, is captured in the VOF simulation. Interestingly, this asymmetry was attributed previously and in a different geometry to thermal effects \citep{Sun09}, while the present calculations ignore thermal effects. Thus, for the present case the asymmetry is a result of viscosity, i.e. the formation of boundary layers. In contrast, the inviscid Rayleigh-Plesset model also plotted in figure \ref{fig:comparison} is symmetric in time.

The late stage of bubble collapse cannot be captured with the VOF model as it assumes that the liquid and bubble content are immiscible, while in experiments the laser generated bubble mainly consists of condensible vapour. Thus, our simulation predicts a milder collapse with re-expansion of the bubble. However, the experiments find a much smaller minimum bubble radius followed by fragmentation of the bubble.

\subsection{Flow reversal due to adverse pressure gradient}\label{subsec:flow_reversal}

\begin{figure}
\begin{center}
\includegraphics[width=0.9\textwidth]{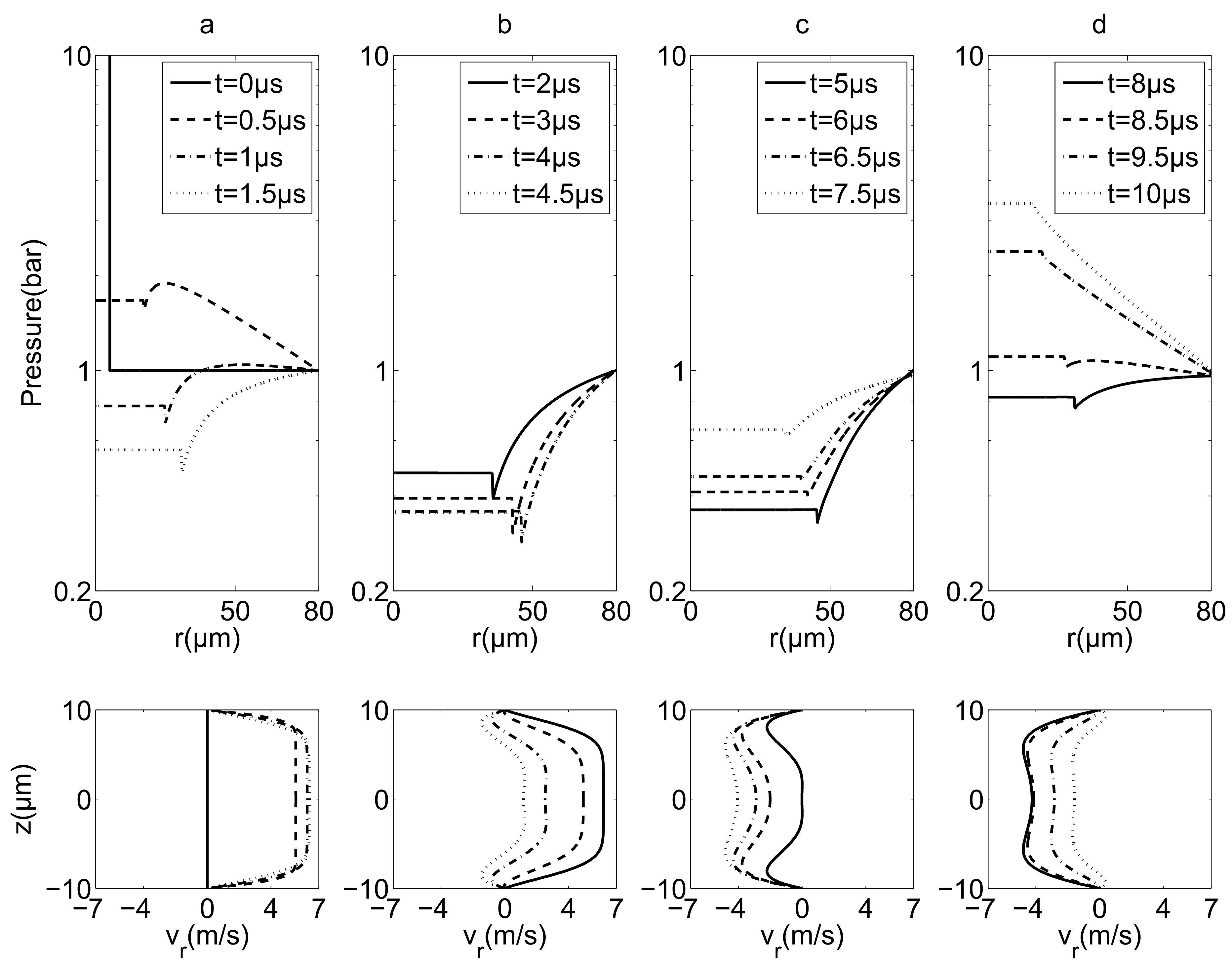}
\end{center}
\caption{Evolution of pressure (top row) and radial velocity (bottom row) during early (a) and late expansion (b) and early (c) and late collapse (d). The times for the pressure and velocity profiles are indicated in the upper legend of each column.}
\label{fig:pressure_evolution}
\end{figure}

The evolution of pressure in the channel and the liquid velocity profile at some distance from the bubble centre are shown in figure \ref{fig:pressure_evolution}; for clarity we have divided the results in four different stages, from left to right: early expansion, late expansion, shrinkage, and rebound. The upper frames in figure \ref{fig:pressure_evolution} show the unsteady pressure in the centre of the channel from $r=0$ to the outlet, i.e. {$r=$\SI{80}{\micro m}}. The axial pressure gradient is negligible in comparison to the significant pressure variation in the radial direction ($\partial p/\partial z \ll \partial p/\partial r$). Therefore, the pressure profile near the wall is highly similar to the profile at the centre, readily observable in pressure field contours in figure \ref{fig:contour_plots}. The gas-liquid interface can be easily identified by the small pressure jump due to surface tension in the pressure profiles in figure \ref{fig:pressure_evolution}. The lower frames of figure \ref{fig:pressure_evolution} show the velocity profile in the liquid at a fixed distance of {$r=$\SI{50}{\micro m}} from the bubble centre.
 
During the first stage, figure \ref{fig:pressure_evolution}a, {$0<t<1.5$\SI{}{\micro s}}, the gas pressure initially at {\SI{100}{bar}} is accelerating the liquid outwards from zero velocity to almost {\SI{7}{m/s}}. In consequence, the gas pressure drops within {\SI{1.5}{\micro s}} below the pressure at the outlet. During this time a flat-top velocity profile develops.

The second stage, figure \ref{fig:pressure_evolution}b, corresponds to the deceleration of the flow to the maximum bubble volume at {$t=$\SI{4.3}{\micro s}}.  As the pressure in the bubble drops, an adverse pressure gradient develops ($\partial p/\partial r>0$). This leads to detachment of the boundary layer and a reversed flow at the boundary sets in. This flow reversal is clearly visible at {$t=$\SI{4}{\micro s}} in the lower frame of figure \ref{fig:pressure_evolution}b. At this stage the flow profile possesses an inflection point; the liquid in the centre of the channel continues to flow towards positive $r$ while at the boundaries the flow is directed towards the bubble. 

In the third stage, figure \ref{fig:pressure_evolution}c, the bubble shrinks, i.e. a net flow towards the bubble sets in and eventually a purely negative velocity profile builds up. In this stage the pressure gradient is stabilising the boundary layer and the pressure in the bubble steadily builds up.

In the last stage the bubble reaches minimum volume and rebounds, figure \ref{fig:pressure_evolution}d, {$8<t<10$\SI{}{\micro s}}. Similar to stage one, the internal bubble pressure is higher than the liquid pressure, but here the liquid flow is toward the bubble. Therefore, the pressure gradient once again opposes the liquid flow and leads to flow reversal near the channel walls. The flow could be described similar to stage two, but with opposite signs (${\partial p}/{\partial r}<0$). The liquid flow is eventually reversed at the walls in {$t=$\SI{10}{\micro s}}, minimum bubble volume is reached, and the bubble begins to rebound.

\subsection{Vorticity generation}
\label{subsec:vorticity_generation}

\begin{figure}
\begin{center}
\includegraphics[width=0.9\textwidth]{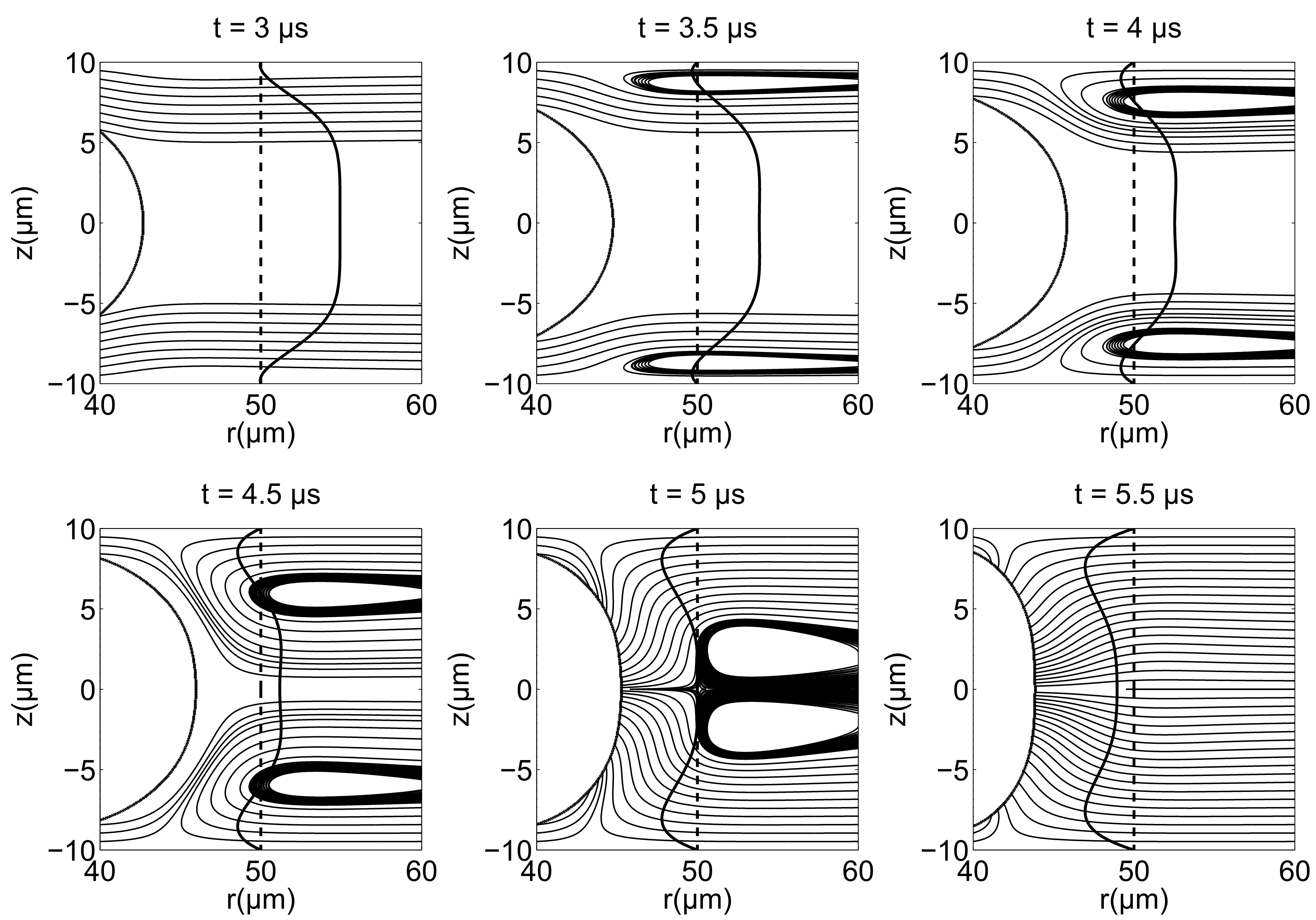}
\end{center}
\caption{Plot of the liquid streamlines during boundary layer separation, i.e. from $t=3$ to \SI{5.5}{\micro s}. The radial velocity profile is plotted at {$r=$\SI{50}{\micro m}}. Two counter rotating vortices form, move upwards and recombine.}
\label{fig:vorticity_generation}
\end{figure}

Figure \ref{fig:vorticity_generation} depicts the instantaneous streamlines during the build up and decay of the adverse pressure gradient together with a radial velocity profile at {$r=$\SI{50}{\micro m}}, corresponding to the late stage of bubble expansion in figure \ref{fig:pressure_evolution}b. At time {$t=$\SI{3.5}{\micro s}} formation of vortices near the channel walls is observed. From $t=3$ to \SI{4.5}{\micro s} the adverse pressure gradient supports the detachment of the boundary layer, leading to a recirculating flow. Due to symmetry two vortex rings are formed. The vortices are transported towards the centre of the channel. In figure \ref{fig:vorticity_generation} the horizontal distance between the vortex core and the bubble wall remains approximately constant at about \SI{5}{\micro m}. Additionally, the separation point of the boundary layer is relatively stable in space, here around {$r=$\SI{44}{\micro m}}, and only moves toward the bubble in the last frame of figure \ref{fig:vorticity_generation}, i.e. at {$t=$\SI{5.5}{\micro s}} when the bubble gains inward speed. 
As both vortex rings migrate towards the centre of the channel they merge, see figure \ref{fig:vorticity_generation}, $t=5.0$ and \SI{5.5}{\micro s}. From then on, the radial flow and the pressure gradient are aligned, stabilising the boundary layer.

\subsection{Flow field contours}
\label{subsec:flow_field_contours}

\begin{figure}
\begin{center}
\includegraphics[width=0.9\textwidth]{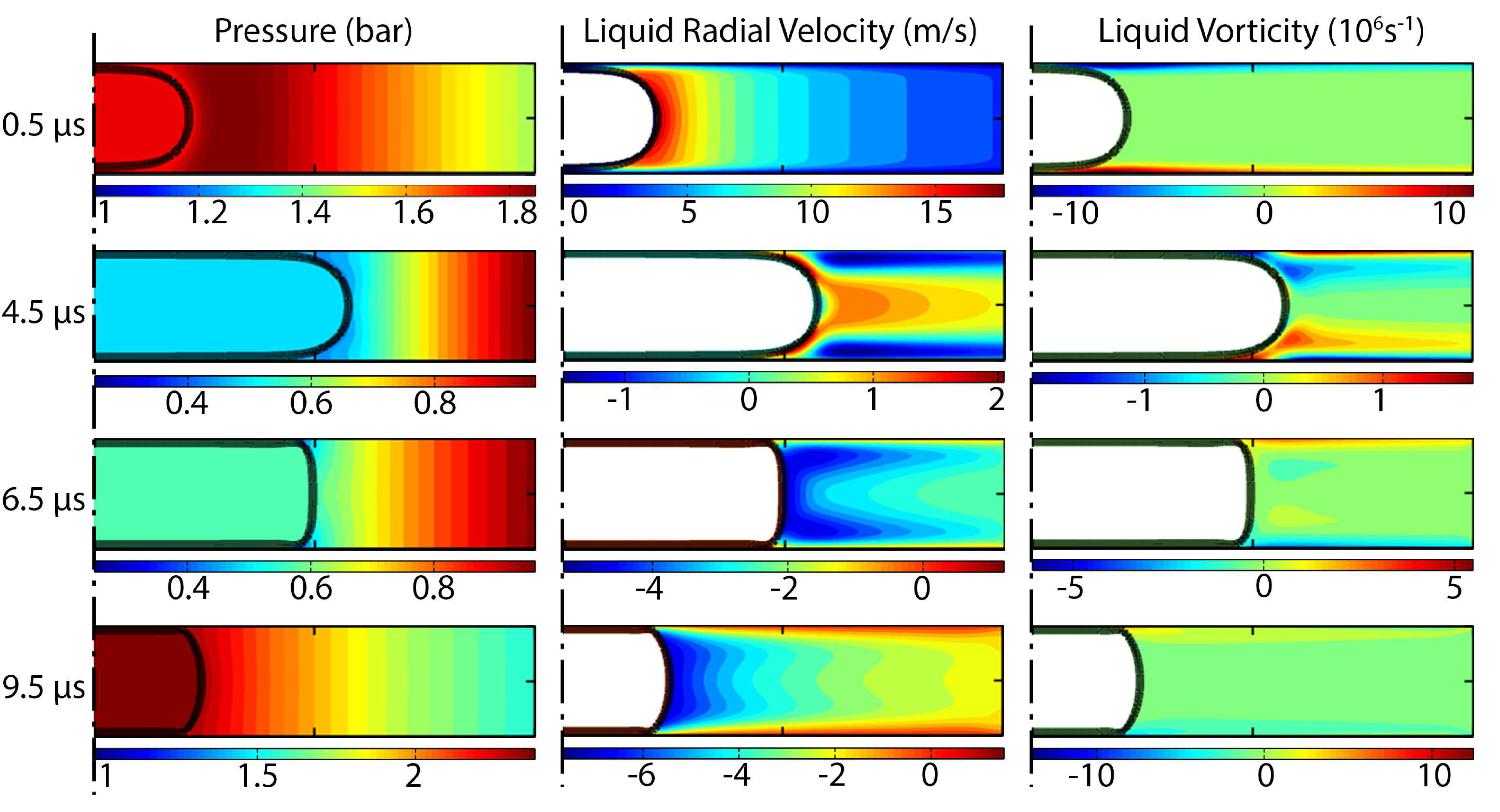}
\end{center}
\caption{Sequence of contour plots of the flow field during bubble expansion and shrinkage. Pressure field in both phases, radial velocity in the liquid, and vorticity in the liquid are shown at left, centre, and right, respectively. The bubble-liquid interface is shown by the black line, i.e. contour of $0.5$ liquid volume fraction. The colour bar for each figure is given underneath. Each figure is plotted with its specific colour map to better distinguish the flow field. Results are shown for a sample time instant of each different stage of the flow.}
\label{fig:contour_plots}
\end{figure}

For the purpose of clarity, we present a sequence of contour plots of the flow field during the bubble expansion and shrinkage cycle. Figure \ref{fig:contour_plots} summarises the simulation results by showing the pressure field in both phases, radial velocity in the liquid, and vorticity in the liquid. Using the same division of results presented in \ref{subsec:flow_reversal}, the sequence of flow field is presented for a sample time instant of each four stage of the bubble pulsation cycle: early expansion ({$t=$\SI{0.5}{\micro s}}), late expansion ({\SI{4.5}{\micro s}}), early shrinkage ({\SI{6.5}{\micro s}}), and rebound ({\SI{9.5}{\micro s}}). 

At the early expansion of the bubble, (e.g. {$t=$\SI{0.5}{\micro s}}) in figure \ref{fig:contour_plots}, extremely high pressure in the bubble accelerates the liquid outward of the channel, causing a Poiseuille-like radial velocity distribution. Though no circulating regions are visible in the flow, there is shear and vorticity in the proximity of channel walls.

The late expansion and early shrinkage stage of the bubble (e.g. {$t=$\SI{4.5}{\micro s}}) is accompanied by the presence of reversed flow and boundary layer separation due to adverse pressure gradient in the liquid. The outward flow is stopped and reversed close to the walls and two counter rotating vortices are observed. The reversed flow and flow circulation in this stage cause significant shear stress near the walls.

As the inward flow sets in during the shrinkage (e.g. {$t=$\SI{6.5}{\micro s}}), the two vortices migrate toward the channel center until they eventually merge and circulation is no more visible in the flow field. The liquid flow toward the bubble is not opposed but stabilised by the pressure gradient.

Finally, the shrinkage causes the pressure to build up in the bubble (e.g. {$t=$\SI{9.5}{\micro s}}), which in turn acts as an adverse pressure gradient, opposing the inward liquid flow, and the bubble rebounds afterwards.

It is worthy to recall that the solution to the Navier-Stokes equations using the volume of fluid method is shared among all phases, i.e. the water vapor in the bubble and its surrounding liquid water. This means that all variables, including the pressure and velocity, are obtained for both phases. In this manuscript, however, we focus only on the liquid flow field induced by the shared pressure field. Therefore in figure \ref{fig:contour_plots},  the radial velocity field and the vorticity field are shown only in the liquid.  

\section{Analytical approximation for liquid velocity profile: pulsating pressure-driven flow}\label{sec:analytical_approximations}

The CFD solution reveals a complex flow pattern, yet we speculate that the main characteristics of the flow can be captured with fundamental solutions of unsteady flows. In the early stage of bubble expansion, the liquid in the channel is accelerated from rest to a velocity profile qualitatively similar to a Poiseuille flow. This acceleration of the liquid is accompanied by a rapid reduction in pressure inside the bubble. 

Except for the short initial period described above, the pressure gradient acting on the liquid has a pulsating behaviour for the majority of the bubble expansion and shrinkage cycle. At the onset of bubble shrinkage, the outlet pressure is higher than the pressure inside the bubble. Therefore, the flow near the boundaries is opposed by an adverse pressure gradient and is eventually reversed. As the inward flow sets in, the bubble shrinkage results in an increase in density and pressure of the gas, which in turn leads to opposition against the inward flow. This situation resembles an oscillating pressure gradient acting on the liquid between two parallel plates. Here we compare the simulation results with an analytical solution for 2-dimensional flow within a gap induced by an oscillating pressure gradient. Although the numerical solution is obtained for an axisymmetric geometry, we simplify the geometry to a planar flow in order to obtain an analytical solution.

\begin{figure}
\begin{center}
\includegraphics[width=0.8\textwidth]{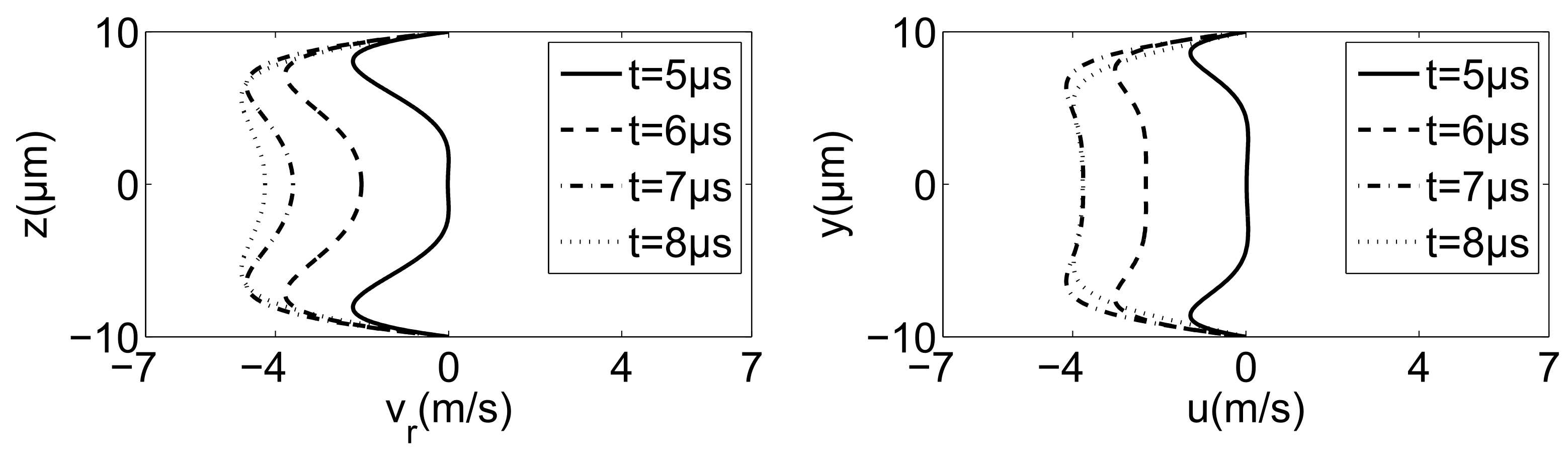}
\end{center}
\caption{Comparison of {\em(left)} the VOF simulation of the radial velocity with {\em(right)} the analytical approximation of the planar velocity profile during bubble shrinkage (\ref{eq:pulsating_pressure_gradient_solution}).}
\label{fig:analytical_comparison_shrinkage}
\end{figure}

Using a harmonic function for the pressure gradient, i.e. $\partial p/\partial x=\Delta p/\Delta x\cdot\sin{\omega t}$, we obtain the transient solution to the 2-dimensional Navier-Stokes equation for incompressible flow between two parallel plates. To be succinct, we do not repeat the  steps here and refer the reader to \citep{pozrikidis2009fluid} for details of obtaining the analytical solution. The solution to this pulsating pressure-driven flow can be expressed as

\begin{equation}\label{eq:pulsating_pressure_gradient_solution}
  u(y,t)=\frac{-1}{\rho \omega} \frac{\Delta p}{\Delta x} \left( f_c(y) \cos{\omega t} + f_s(y) \sin{\omega t} \right) \quad,
\end{equation}

\noindent where $f(y)$ is given as a complex function with real and imaginary parts, $f_c(y)$ and $f_s(y)$, respectively. The complex function $f(y)$ is formulated as

\begin{equation}\label{eq:f(y)}
  f(y)=f_c(y)+if_s(y)=1-\frac{\cosh{\left( (\frac{2y}{h}-1) \frac{h}{2} \sqrt{\frac{-i \omega \rho}{\mu}} \right)}}{\cosh{\left( \frac{h}{2} \sqrt{\frac{-i \omega \rho}{\mu}} \right)}} \quad.
\end{equation}

The pulsating velocity profile for the bubble shrinkage is approximated with the solution from (\ref{eq:pulsating_pressure_gradient_solution}) and is compared with the CFD solution in figure \ref{fig:analytical_comparison_shrinkage}. A value of ${\Delta p}/{\Delta x}=2.5 \cdot 10^4$ bar$\cdot$m$^{-1}$ with a period of $2\pi/\omega=$\SI{10}{\micro s}  is used for the harmonic pressure gradient. This simplified analytical description is able to capture the general features of the bubble-driven flow during bubble shrinkage. Development of a reversed flow near the channel walls is observed, which later on reverts completely toward the bubble. Later on, the pulsating pressure gradient opposes and decelerates the inward flow, which is similar to the findings in the numerical solution. 

\section{Experiments: deformation of cells}
\label{experimental_study}

\begin{figure}
\begin{center}
\includegraphics[width=0.8\textwidth]{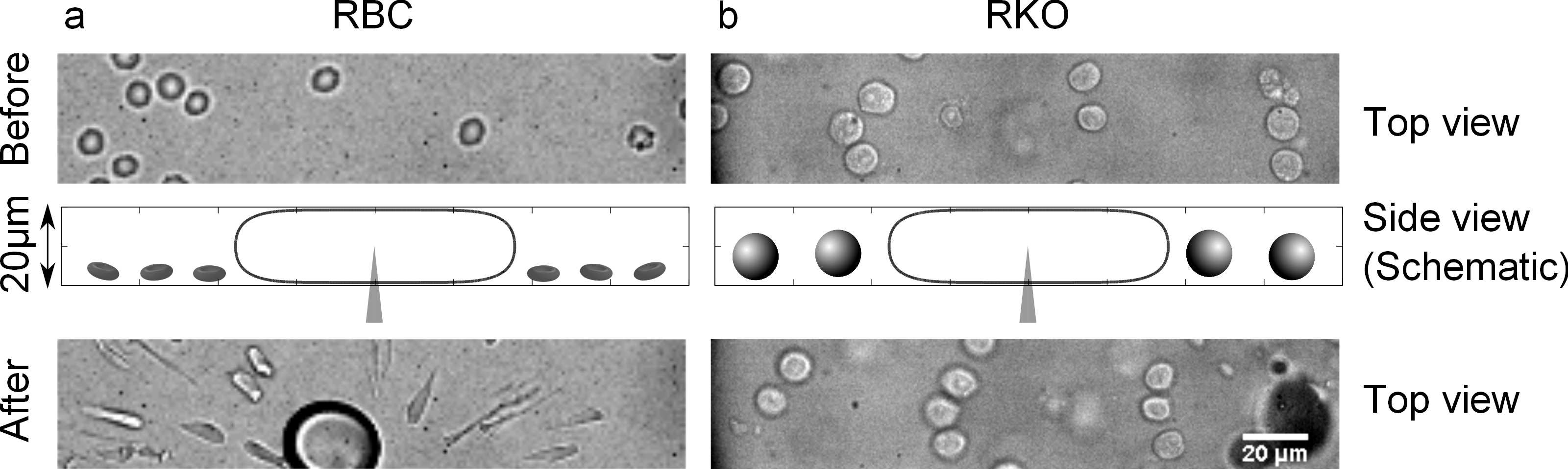}
\end{center}
\caption{Deformation of two different cell types due to the cavitation bubble induced flow in a \SI{20}{\micro m} thin gap. a) The top shows red blood cells (RBCs) just before a laser creates a single cavitation bubble and at the bottom are the cells just after the bubble collapse. b) Colon cancer cells (RKOs) exposed to a similar flow before and after the cavitation bubble dynamics. RKOs are much less deformed compared to RBCs (see text).}
\label{fig:experimental_results}
\end{figure}

The numerical simulation suggests that deformable objects located within the boundary layer and close to the bubble may be stretched considerably by the shearing flow. This was indeed observed for red blood cells in previous experimental studies using a microfluidic gap \citep{quinto2011red, li2013yield}. Red blood cells are thin, bi-concave cells with a diameter of about \SI{8}{\micro m} and a thickness of less than \SI{2}{\micro m}. After they are placed inside the gap they sediment to the bottom because of their higher density than the surrounding fluid (saline buffer solution). Figure \ref{fig:experimental_results}a is showing a typical configuration of red blood cells (RBCs) shortly before (top) and after the bubble oscillation (bottom) viewed from the top. RBCs near the bubble are largely stretched. Details of the experiment are available in \citep{quinto2011red, li2013yield}.

Interestingly, larger cells show very little deformation under similar flow condition, such as colon cancer cells (RKO), see figure \ref{fig:experimental_results}b. The main difference is that RKO cells have a spherical shape with a diameter of $10$ to \SI{15}{\micro m} and therefore span a large part of the microfluidic gap. We now discuss a possible explanation of the marked difference.

Since the thickness of RBCs is less than \SI{2}{\micro m} they are located close to the bottom channel wall and are of comparable size as the vortex. After being initially translated away from the bubble center, during the late stage of expansion and the onset of shrinkage the RBCs are exposed to a shearing force from the vortices near the wall. From the simulation results, neglecting the presence of cells, we obtain values of about \SI{2e6}{1/s} for the magnitude of the shear strain rate, $1/2 (\partial v_r/\partial z + \partial v_z/\partial r)$, near the walls, where the RBCs are located. This can be related to a stretching of a fluid particle assuming a characteristic height and duration. Inserting \SI{2}{\micro m} and a duration of \SI{2}{\micro s} we obtain a length increase of \SI{8}{\micro m}, a value which is in the order of the observed stretching in experiments, e.g. \citep{li2013yield}.

To the contrary, the cells extending into the centre of the channel, such as the RKO cells, will be advected with the flow. The shear stresses outside the boundary layer are considerably lower and therefore lead to weaker deformation. We find about 10\% strain at the centre of the channel over a thickness of \SI{1}{\micro m}. The resulting deformation of the cell may be well within the resolution limit of the imaging optics.

\section{Discussion}
\label{sec:discussion}

A notable feature of the present numerical solution to the flow induced by a cavitation bubble in a narrow gap is the presence of boundary layer separation, reversed flow, and recirculation. In similar geometries, flow reversal has been reported by other investigators. In the case of {\em steady}, radial flow between two flat disks, extension of \citet{von1999note} on the power-series solution obtained in \citep{zitouni1997purely} shows the existence of boundary layer separation in radial diffusers. Prior to that, flow visualisation done in \citep{mochizuki1985self} on radial flows with a steady influx showed the nucleation, growth, migration, and eventual decay of vortices in the outward flow. \citet{mochizuki1985self} also obtained finite-difference solutions to the unsteady vorticity transport equation which was in agreement with their experimental findings. In an analytical treatment of the unsteady axisymmetric flow between two flat disks, \citet{elkouh1975oscillating} found the radial velocity distribution induced by an oscillating source/sink exhibits flow reversal near the walls.

In a different confining geometry, an expanding bubble in a tube, direct numerical simulation of the flow reveals the presence of a recirculation region between the wall and the core flow at the end of bubble growth and beginning of shrinkage \citep{ye2004direct}. However, \citet{ye2004direct} do not discuss this flow feature in greater depth, and instead focus on the wall pressure and shear stress which is aligned with the main application of their study, gas embolotherapy. Our findings about the wall shear stress due to bubble induced liquid flow in a microfluidic geometry agrees well with their report; thus, we refer the reader to \citep{ye2004direct} for a detailed discussion about wall shear stress. Interestingly, even inviscid boundary layer simulations may explain the deformation of dispersed droplets (being a simple model for a cell) if their interfacial tension is accounted for, see \citep{tandiono2013resonant}. They simulated a droplet at some distance from an oscillating bubble in an infinite liquid. The interfacial tension introduces to second timescale besides the period of bubble oscillation, i.e. it causes the droplet to oscillate in a surface mode.

Utilising a viscous model in the present study, we find that during the late expansion of a cavitation bubble in a narrow axisymmetric gap, the adverse pressure gradient leads to boundary layer separation and flow reversal in the proximity of the walls. Although such complex flow patterns in axisymmetry require computational treatment of the problem, a simplified analytical expression for planar pulsating flow supports our understanding of the series of events we observe in the numerical simulations and experiments. A prominent result of the observed flow reversal and vorticity is generation of strong shear stress in the liquid close to the boundaries. The results indicate that some flows created by oscillating bubbles in narrow gaps may not be accurately described with inviscid potential flow in cylindrical symmetry. Particularly, in studying the mixing/emulsification of flows or the deformation of elastic objects, such as yeast cells \citep{tandiono2012sonolysis}, it is necessary to account for the vorticity generation and transport. For both applications, it would be interesting to extend the present simulation with two-way coupling. 

\bibliography{arxiv-milad-references}

\end{document}